\documentclass[12pt]{article}

\usepackage{amssymb,amsmath,amsthm,epsfig}

\theoremstyle{plain}
\newtheorem{lem}{Lemma}[section]
\newtheorem{thm}[lem]{Theorem}
\newtheorem{cor}[lem]{Corollary}

\theoremstyle{definition}
\newtheorem{defn}{Definition}[section]

\theoremstyle{remark}
\newtheorem{rem}{Remark}[section]

\newcommand{\n}{\noindent}
\newcommand{\ds}{\displaystyle}

\begin{document}
\title{Quantum Multi-object Search Algorithm with the\\ Availability 
 of Partial Information}
\author{Goong Chen$^*$ and Zijian Diao$^*$} 

\date{}
\maketitle

\begin{center}{\bf ABSTRACT}
\end{center}\smallskip

Consider the unstructured search of an unknown number $l$ of items in a large 
unsorted database of size $N$.  The multi-object quantum search algorithm 
consists of two parts.  The first part of the algorithm is to generalize 
Grover's
single-object search algorithm to the multi-object case
([\ref{Bo}, \ref{Br}, \ref{Ch1}, \ref{Ch2}, \ref{Ch3}]) and the second 
part is to solve a counting problem to determine $l$ 
([\ref{Br}, \ref{Mo}]).  In this paper, we study the multi-object
 quantum search algorithm
(in continuous time), but in a more structured way by taking 
into account the availability of partial
information.  The modeling of available partial information is
done simply by the combination of several prescribed, possibly
overlapping, information sets with varying weights to signify the reliability 
of each set. The associated statistics is estimated and the algorithm 
efficiency and complexity are analyzed.

Our analysis shows that the search algorithm described here may not be 
more efficient than the unstructured (generalized) multi-object Grover
search if there is ``misplaced confidence''.  However, if the information
sets have a ``basic confidence'' property in the sense that each 
information set contains at least one search item, then a quadratic
speedup holds on a much smaller data space, which further expedites the
quantum search for the first item.
\vfil
\begin{itemize}
\item[*] Department of Mathematics, Texas A\&M University, College Station,
TX \ 77843-3368.\newline
E-mails: \ gchen@math.tamu.edu, zijian.diao@math.tamu.edu.
\end{itemize}
\newpage

\section{Introduction}\label{sec1}

\indent
Grover's quantum search algorithm, since its first publication 
in 1996 ([\ref{Gr1}]), has become one of the most prominent algorithms 
in quantum computation.  Its elegance has drawn the attention of numerous 
computer scientists, mathematicians and physicists, resulting in many research 
papers on this subject.  Grover's original work 
[\ref{Gr1}, \ref{Gr2}, \ref{Gr3}] dealt with a single-object
search in a large unsorted database.  He shows that his quantum algorithm
has a quadratic speedup.  Farhi and Gutmann [\ref{Fa}] 
presents a continuous time, or ``analog analogue'' version, of Grover's 
algorithm and obtains a similar complexity.

In practice, most of the search tasks consist of finding more than one item 
in a large database.  Therefore the development of multi-object search 
algorithms
is important.  By utilizing the two most important ingredients in Grover's
algorithm, namely,

\begin{itemize}
\item[(i)]  the notion of amplitude amplification; and
\item[(ii)]  the dramatic reduction to invariant subspaces of low dimension
for the unitary operators involved,
\end{itemize}
it is possible to generalize the algorithm to multi-object search.  See
the discrete-time case in Boyer, Brassard, H\o yer and Tapp [\ref{Bo}], and the 
continuous-time case in Chen, Fulling and Chen [\ref{Ch2}].  However, for 
multi-object problems, the number of search items is normally not given 
a priori and,
therefore, its determination is crucial.  This becomes a 
\emph{quantum counting problem}.  
The problem was partly treated in Brassard, H\o yer and Tapp
[\ref{Br}] but a complete solution did not seen to appear until Mosca's Ph.D.
Thesis [\ref{Mo}] in 1999.  The counting problem can be studied with the 
techniques of ``eigenvalue kickback'', phase/amplitude estimations and 
quantum Fourier transforms (QFT).

Excluding the computational complexity of the counting problem, the 
generalized, \emph{unstructured} Grover multi-object search of 
$l$ items in a database of $N$ items
has computational complexity $O(\sqrt{N/l})$ versus the classical
$\Theta(N/(l+1))$ ([\ref{Mo}, p.70]).  So again we see a quadratic speedup. 
 This is
significant.  Nevertheless, pragmatically, one usually can (and should) do
much better than this because in most realistic search tasks there is
additionally given partial information about the search targets, provided
that one knows how to utilize such information.  

The mathematical modeling of the availability of partial information
is challenging work.  Obviously, there are varied situations of how
such information can be given and how it can be encoded into the computer.
Therefore, mathematical expressions intended to model those situations
may be qualitatively different.  This difficulty is further compounded by 
the fact that no quantum computers (QC) have been built and are currently
in operation so far, as solutions to the modeling problem hinges very much 
on the addressing, retrieval and data structure designs of the future QC.  At
present, we do not yet know how to categorize all (or most) of the
possible situations that may naturally arise, but we are continuing to
probe in this direction to improve our understanding on this modeling
aspect.  Our work here, though rather simplistic in nature, hopefully could
serve as a modest start to draw more research interest in the directions
of \emph{structured search} in the future.

Consider the following hypothetical situation:

\begin{eqnarray}\label{eq1}
&&
\textrm{``Professor John Smith, an outdoors buff, goes to the libarary.  He } 
\nonumber\\
&& \textrm{requests the librarian to assist him to find the total 
number and} \nonumber\\
&& \textrm{the titles of the books published between 1/15/1990 and 
6/15/1990}\nonumber\\
&& \textrm{on the subjects of hunting, fishing or hiking''.}
\end{eqnarray}

His search targets are precisely given as follows:
\begin{eqnarray}\label{eq2}
\mathcal{T}&=&\{ \textrm{book title } x| x \textrm{ is published between 
1/15/1990 and 6/15/1990, } \nonumber\\ 
&&  x \textrm{ is on hunting, fishing or hiking}\}.
\end{eqnarray}

The number of items in $\mathcal{T}$ is not known in advance; therefore, it 
involves a counting problem as well.  A brute force multi-object (generalized)
Grover search would proceed to find items in $\mathcal{T}$ among all books 
in the library's holding, denoted as $\bar{A}$.  This would require the crude
$O(\sqrt{N/l})$ quantum complexity if $\mathcal{T}$ has cardinality $l$ and the
library's book holding $\bar{A}$ has cardinality $N$.  This would be
inefficient.  However, (most) libraries group books according to 
subject interests.  Instead of searching $\mathcal{T}$ among $\bar{A}$, we 
should search $\mathcal{T}$ among $A_1 \cup A_2 \cup A_3$, where $A_1$, $A_2$ 
and $A_3$ denote, respectively, the set of book titles on hunting, fishing
and hiking.  This is intuitively clear to surely cut down search time
even without mathematical justifications first.  See (I2) in \S 3.

We call such sets $A_1$, $A_2$ and $A_3$ here (partial) information
sets.  These sets may not be disjoint from each other, such as example
(1.1) here amply illustrates the fact that there are many books dealing
with \emph{both} hunting and fishing and, thus, they belong to 
$A_1 \cap A_2$.  Inside a computer (whether quantum or electronic), 
each of such datasets
like $A_i$, $i=1, 2, 3$, here occupies a block of memory space, with additional
ordered/sorted data structure.  For example, the dataset $A_1$ containing
all book titles on hunting may already be either sorted according
to the alphabetical orders of authors' names or the chronological
orders of time of publication, or both.  Such ordered data structures
are likely to even expedite search with possible 
\emph{exponential speedup};
nevertheless, we will not consider or exploit any sorted data structure 
for the time being in this paper.

Generally, for a given collection of information sets $A_i$, $i=1,2,\ldots, n$,
such that $\mathcal{T}\subseteq A_1 \cup A_2 \cup \ldots \cup A_n$, there is in 
addition a given probability distribution that weighs some sets $A_j$
more heavily than the others, depending on the reliability or preferences of
the information source.  For example, in (1.1), if Professor Smith has
indicated that fishing is his primary sporting interest, then his information 
set $A_2$ ought to weigh heavier than $A_1$ or $A_3$ in his case.

Now having offered the physical motivations in our study of the
modeling of search with the availability of partial information, we 
proceed to treat the multi-object search problem related to an 
analogue QC design.

\section{Multi-Object Search with the Availability of Partial Information
on an Analogue Quantum Computer}

\indent

\setcounter{equation}{0}
Let a large database consist of $N$ unsorted objects $\{ w_j| 1\le j \le N\}
\equiv \bar{A}$, where $N$ is an extremely large integer.  Let 
$\mathcal{T}\equiv \{ w_j| 1\le j\le l\} \subset \bar{A}$ be the target set 
of search objects, where $l$ is an unknown integer.  The information about 
$\mathcal{T}$ is given as follows:

\begin{itemize}
\item[(1)]  There is an oracle (or Boolean) function satisfying
\begin{equation}\label{eq4}
f(w_j) = \left\{\begin{array}{ll}
1,&j=1,2,\ldots, \ell,\\
0,&j=\ell +1, \ell+2,\ldots, N.\end{array}\right.
\end{equation}
This function acts in the black box of QC and can be known only through
queries.
\item[(2)]  There are $n$ explicitly given information (sub)sets $A_j$, 
$j=1, 2, \ldots, n$, such that 
\[
 A_j= \{ \underset{\sim}{w}{}_{j, i} | i=1, 2, \ldots, k_j\} \subset \bar{A}
\]
and 
\begin{equation}
\mathcal{T}\subseteq A_1 \cup A_2 \cup \ldots \cup A_n
\end{equation}
\item[(3)]  There is a given probability distribution that assigns different
weights to various subsets $A_j$, depending on the reliability or (searcher's)
preference of that information set.  Let such weights be called 
\emph{reliability coefficients} and denoted as
\begin{equation}
\{\alpha_j >0| j=1, 2, \ldots, n, \sum_{j=1}^n \alpha_j=1 \}
\end{equation}
\end{itemize}

In the QC, each object $w_j \in \bar{A}$ is stored as an 
eigenstate $|w_j\rangle$
which collectively form an orthonormal basis $B\equiv \{|w_j\rangle |
j=1, 2, \ldots, N\}$ of an $N$-dimensional Hilbert space $\mathcal{H}$.  
Let us denote $\mathcal{L}= span\{|w_j\rangle|j=1, 2, \ldots, l\}$ 
as the subspace 
containing all the eigenstates representing the search targets.  Suppose 
we are given a Hamiltonian $ \widetilde{H}$ in $\mathcal{H}$ and we are 
told that $ \widetilde{H}$ has an eigenvalue $E\ne 0$ on the entire 
subspace $\mathcal{L}$ and all the other eigenvalues are zero.  
The search task is 
to find an eigenstate $|w_j\rangle$ in $\mathcal{L}$ that has 
eigenvalue $E$.  The 
task for the first search item is regarded as complete when a measurement 
of the system shows that it is in a state $|w_j\rangle \in \mathcal{L}$.

The analogue quantum computer for implementing multi-object
Grover's search is a quantum process modeled by the Schr\"odinger
equation
\begin{equation}\label{eq6}
\left\{\begin{array}{ll}
i\ds\frac{d}{dt} |\psi(t)\rangle = H|\psi(t)\rangle,&t>0,\\
|\psi(0)\rangle = |s\rangle,\end{array}\right.
\end{equation}
where $H$, the overall Hamiltonian, is given by
\begin{align}
H= \widetilde{H} +H_D,
\end{align}
where 
\begin{align}
 \widetilde{H} = E\sum_{j=1}^l |w_j\rangle\langle w_j|
\end{align}
is the Hamiltonian satisfying the aforementioned property that it has an
eigenvalue $E$ on $\mathcal{L}$, 
with the rest of its eigenvalues being zero.  Note
that 
\[
 \widetilde{H}= \frac{E}4 \sum^N_{i=1} [|w_i\rangle - (-1)^{f(w_i)}|w_i\rangle]
[\langle w_i| - (-1)^{f(w_i)} \langle w_i|];
\]
therefore the knowledge of $f$ alone determines $ \widetilde{H}$; no 
knowledge of $\{ |w_j\rangle | 1\le j\le l\}$ is required or utilized since it
is assumed to be hidden in the oracle (black box).

In (2.5), $H_D$ is the ``driving Hamiltonian''; its choice is up to the
algorithm designer.

\begin{rem}  Without the assumption (2.2) and (2.3), a ``good'' driving
Hamiltonian to choose ([\ref{Fa}, \ref{Ch2}]) is
\begin{equation}
H_D = E|s\rangle \langle s|
\end{equation}
related to the initial state $|s\rangle$, where $|s\rangle$ is further 
chosen to be
\begin{equation}
|s\rangle = \frac1{\sqrt N} \sum^N_{j=1} |w_i\rangle,
\end{equation}
the uniform superposition of all eigenstates.

For the discrete-time case ([\ref{Bo}, \ref{Ch1}]), the generalized Grover 
``search engine'' is chosen to be 
\begin{equation}\label{eq10a}
U = -I_sI_L,
\end{equation}
where
\begin{align}\label{eq10b}
I_L &= \pmb{I} - \frac{2}{E} \widetilde{H}, 
I_s =\pmb{I} - 2|s\rangle \langle s|,\\
\pmb{I} &= 
\text{the identity operator on the Hilbert space } \mathcal{H}.\nonumber
\end{align}
\end{rem}
$\hfill\square$

Since now we have the extra properties (2.2) and (2.3) at hand, based on the
insights we have gained from the analysis of Grover's algorithm, it is not
difficult to see that searching by using the initial state (2.8) is not 
necessary, because the useful component, namely, the projection of $|s\rangle$ 
in $\mathcal{L}$, is too small compared with the component of 
$|s\rangle$ outside $\mathcal{L}$:
\[
\big\| P_\mathcal{L}(|s\rangle)\big\|^2/\big\| P_{\mathcal{L}^\perp}(|s\rangle)\big\|^2 = l/(N-l),
\]
where $P_\mathcal{L}$ is the orthogonal projection operator onto the subspace 
$\mathcal{L}$, 
$\mathcal{L}^\perp$  is the orthogonal complement of $\mathcal{L}$, and $\| . \|$ is 
the norm of $\mathcal{H}$.

Because of (2.2) and (2.3), instead of (2.8) it is now natural for us to
choose

\begin{align}
|s\rangle &= \frac1\nu \sum^n_{j=1} \sum^{k_j}_{i=1}
\alpha_j|\underset{\sim}{w}{}_{j,i}\rangle
\end{align}
where $\nu$ is a normalization constant.  From (2.11), we rearrange terms
and simplify, obtaining
\begin{align}
|s\rangle=\sum^\ell_{i=1}\beta_i|w_i\rangle+\sum^{\ell+R}_{i=\ell+1}\beta_i
|\underset{\sim}{w}{}_i\rangle,
\end{align}
where the first sum on the RHS above is composed of all the terms in 
$\mathcal{L}$,
and the second sum consists of the remaining R terms in $\mathcal{L}^\perp$.

\begin{rem}
With the choice of a different $|s\rangle$ as in (2.12), the state 
equation (2.4) now has a new initial condition which is different from the
uniform superposition of all eigenstates given in (2.8).  Biron, Biham, 
et al. [\ref{Bi1}, \ref{Bi2}] call this the choice of ``arbitrary initial
amplitude distribution'' in their paper.  The papers [\ref{Bi1}, \ref{Bi2}]
have shown certain advantages of the choice of general amplitudes in the 
discrete time case even though their ideas are unrelated to our problem
under treatment here.$\hfill\square$
\end{rem}

\begin{thm}\label{thm1} 
Consider the Schr\"odinger
equation
\begin{equation}\label{eq6}
\left\{\begin{array}{ll}
i\ds\frac{d}{dt} |\psi(t)\rangle = H|\psi(t)\rangle
=( \widetilde{H}+H_D)|\psi(t)\rangle,&t>0,\\
|\psi(0)\rangle = |s\rangle,\end{array}\right.
\end{equation}
where $ \widetilde{H}$ and $H_D$ are given, respectively, by (2.6) and (2.7), 
and $|s\rangle$ is given by (2.12).  Then
\begin{itemize}
\item[(1)] $H$ and the evolution operator $e^{-iHt}$ have an invariant
two-dimensional subspace\hfil\break $\mathcal{V} \equiv
\text{span}\{|\widetilde w\rangle, |r\rangle\}$, with

\begin{equation}\label{eq12d}
y\equiv\left( \sum^\ell_{i=1} |\beta_i|^2\right)^{1/2} \le 1, ~~|\widetilde
w\rangle \equiv \frac1y \sum^\ell_{i=1} \beta_i |w_i\rangle, ~~|r\rangle 
\equiv \frac1{\sqrt{1-y^2}} \sum^{\ell+R}_{i=\ell+1}
\beta_i|\underset{\sim}{w}{}_i\rangle, 
\end{equation}
On $\mathcal{V}$, $H$ and $e^{-iHt}$ admit $2\times 2$ matrix representations
\begin{align}
\label{eq12e}
H &= E\left[\begin{matrix} 1+y^2&y\sqrt{1-y^2}\\
y\sqrt{1-y^2}&1-y^2\end{matrix} \right],\\
\label{eq12f}
e^{-iHt} &= e^{-iEt} \left[\begin{matrix} 
\cos(Eyt)-iy \sin (Eyt)&-\sqrt{1-y^2} i \sin (Eyt)\\
-\sqrt{1-y^2} i\sin(Eyt)&\cos(Eyt) + iy\sin (Eyt)\end{matrix}\right].
\end{align}
\item[(2)] The state $\psi(t)$ is given by
\begin{equation}\label{eq12c}
\psi(t) = e^{-iHt} |s\rangle = e^{-iEt} \{[y\cos(Eyt) - i\sin (Eyt)]
|\widetilde w\rangle + \sqrt{1-y^2} \cos (Eyt)|r\rangle\}, t>0.
\end{equation}
\end{itemize}
\end{thm}
{\bf Proof:} From (2.12) and (2.14), we have 
\begin{align}
|s\rangle=y |\tilde{w}\rangle +\sqrt{1-y^2}|r\rangle;
\end{align}
so
\[ 
|s\rangle\langle s|= y^2|\tilde{w}\rangle\langle\tilde{w}| 
+y\sqrt{1-y^2}(|\tilde{w}\rangle\langle r|+
|r\rangle\langle\tilde{w}|)+(1-y^2)|r\rangle\langle r|.
\]
Also, note that
\[
 \widetilde{H} = E\sum_{j=1}^l |w_j\rangle\langle w_j| =E P_\mathcal{L}
\]
For any vector $v \in \mathcal{V}$, we may use the spinor notation
\[
v=a|\tilde{w}\rangle +b |r\rangle = [a \quad b]^T; a, b \in \mathbb{C}
\]
Thus, 
\begin{align}
&Hv= ( \widetilde{H} + E|s\rangle\langle s|)v    \nonumber\\
  &=E\Big(P_\mathcal{L}+[y^2|\tilde{w}\rangle\langle\tilde{w}|+
  y\sqrt{1-y^2}(|\tilde{w}\rangle\langle r|+
  |r\rangle\langle\tilde{w}|)+(1-y^2)|r\rangle\langle r|]\Big)
  (a|\tilde{w}\rangle+b|r\rangle)               \nonumber\\
  &=(a|\tilde{w}\rangle+ay^2|\tilde{w}\rangle+ay\sqrt{1-y^2}|r\rangle)+
  (by\sqrt{1-y^2}|\tilde{w}\rangle +b(1-y^2)|r\rangle        \nonumber\\
  &=\left[\begin{matrix} 1+y^2&y\sqrt{1-y^2}\\
   y\sqrt{1-y^2}&1-y^2\end{matrix} \right] 
   \left[ \begin{matrix} a \\ b \end{matrix} \right]\in \mathcal{V}.
\end{align}
Obviously, $H$ is invertible on $\mathcal{V}$.  Therefore, $H(\mathcal{V})
=\mathcal{V}$, and $H$ has the $2 \times 2$ matrix representation (2.15) on
$\mathcal{V}$ according to (2.19).  From (2.15), we calculate the 
exponential matrix $e^{-iHt}$ to obtain (2.16).

The solution (2.17) for the state equation (2.13) follows from (2.17) and
(2.18).  
$\hfill\square$

\begin{cor} 
Assume the same conditions as Theorem 2.1.  Then at time $T=\frac{\pi}{2Ey}$,
we have $|\psi(T)\rangle \in \mathcal{L}$.  Consequently, after 
measurement it yields a
first search item $w_j \in \mathcal{T}$ with probability $\beta_j^2/y^2$, for 
$j=1, 2, \ldots, l$, and total probability 1.
\end{cor}
{\bf Proof:}  Obvious from (2.17).  $\hfill\square$

\begin{thm} 
Assume the same conditions as Theorem 2.1. Define the following two vectors
in $\mathcal{V}$:
\begin{equation}
X_1 = \frac1{\sqrt 2} \left[\begin{matrix}
\sqrt{1+y}\\ \sqrt{1-y}\end{matrix} \right], \quad X_2 =
\frac1{\sqrt 2} \left[\begin{matrix} -\sqrt{1-y}\\
\sqrt{1+y}\end{matrix}\right].
\end{equation}
Then
\begin{itemize}
\item[(i)]  $X_1$ and $X_2$ are the unique orthonormal eigenvectors of $H$ on
$\mathcal{V}$, i.e., (2.15), corresponding, respectively, to eigenvalues
$\lambda_1 = E(1+y)$ and $\lambda_2 = E(1-y)$;
\item[(ii)]  For each $ t\ge 0$, the evolutionary operator $e^{-iHt}$ 
satisfies
\begin{equation}\label{eq12g}
e^{-iHt} X_1 = e^{-iE(1+y)t} X_1,\quad e^{-iHt} X_2 = e^{-iE(1-y)t} X_2.
\end{equation}
\end{itemize}
\end{thm}
{\bf Proof:} Straightforward calculations and verification. $\hfill\square$

Even though Cor. 2.2 gives the informed answer that the quantum search process
should be measured at time $T= \pi/(2Ey)$ in order to obtain the first
desired object, the trouble is that we do not know explicitly what the value
of $y$ is in order to determine $T$.  Now Thm. 2.3 affords the 
information that $X_1$ and $X_2$ are eigenvectors of $H$ of $e^{-iHt}$. 
We can apply the ``eigenvalue kickback'' and ``phase estimation''
techniques, first developed by Kitaev [\ref{Ki}], to estimate the crucial value
of $y$.  The quantum Fourier transforms (QFT) plays a central role in this
approach;  see a lucid introduction in Mosca [\ref{Mo}].

Let us construct a unitary operator $Q\equiv e^{-iH(2\pi/E)}$.  Then
from (2.21) and (2.18), we have

\begin{align}
\label{eq12h}
QX_1 &= e^{-i2\pi y} X_1, \quad QX_2 = e^{i2\pi y} X_2,\\
Q^m |s\rangle &= Q^m(y|\widetilde w\rangle + \sqrt{1-y^2} |r\rangle) = Q^m
\left(\sqrt{\frac{1+y}2} X_1 + \sqrt{\frac{1-y}2}X_2\right)\nonumber\\
\label{eq12i}
&= \sqrt{\frac{1+y}2} e^{-i2m\pi y} X_1 + \sqrt{\frac{1-y}2} e^{i2m\pi y}
X_2, \text{ for } m=0,1,2,\ldots~.
\end{align}

Thus we see that $y$ appears as a phase factor in (2.22) and (2.23).
Further, $y$ also appears in the amplitudes on the RHS of (2.23).

We add an
ancilla register $|m\rangle$, $m=0,1,2,\ldots, M-1$, for a sufficiently large
integer $M$ and form 
\begin{equation}\label{eq12j}
|\Psi_1\rangle \equiv \sum^{M-1}_{m=0} |m\rangle \otimes Q^m |s\rangle =
\sqrt{\frac{1-y}2} \sum^{M-1}_{m=0} e^{i2m\pi y} |m\rangle \otimes X_2 +
\sqrt{\frac{1+y}2} \sum^{M-1}_{m=0} e^{i2m\pi(1-y)} |m\rangle \otimes X_1.
\end{equation}
For any given $|x\rangle$, $x=0,1,\ldots, M-1$, define QFTs $\mathcal{F}_M$
and $\mathcal{F}^{-1}_M$ by
$$\mathcal{F}_M|x\rangle = \frac1{\sqrt M} \sum^{M-1}_{k=0} e^{i2k\pi x/M}
|k\rangle,\quad \mathcal{F}^{-1}_M |x\rangle = \frac1{\sqrt M}
\sum^{M-1}_{k=0} e^{-i2k\pi x/M} |k\rangle.$$
For any $\omega\in\mathbb{R}$, define
\begin{align}
|\widetilde\omega\rangle = \mathcal{F}^{-1}_M \left(\frac1{\sqrt M}
\sum^{M-1}_{k=0} e^{i2k\pi \omega} |k\rangle\right).
\end{align}
Applying $\mathcal{F}^{-1}_M$ to the first register in (2.24), we
obtain
\begin{equation}\label{eq12k}
|\Psi_2\rangle \equiv \sqrt{\frac{1-y}2} |\widetilde y\rangle \otimes X_2 +
\sqrt{\frac{1+y}2} |\widetilde{1-y}\rangle \otimes X_1.
\end{equation}
Now, measurement of the first register on the RHS of (2.26) will yield
the state $|\widetilde y\rangle$ or $|\widetilde{1-y}\rangle$ with probability
$\frac{1-y}2$ and $\frac{1+y}2$, respectively. The state $|\widetilde
y\rangle$ or $|\widetilde{1-y}\rangle$ further collapses to one of the
eigenstates $|\pmb{j}\rangle$, $\pmb{j} = 0,1,2,\ldots, M-1$, of the first
register.

\begin{thm}
Assume the same conditions as Theorem 2.1.  Let us measure the first register
of $|\Psi_2\rangle$ on the RHS of (2.26), which collapses to one of the 
eigenstates
$|\pmb{j}\rangle$, $\pmb{j}=0, 1,2, \ldots, M-1$, of the first register.  Then

\begin{align}
\label{eq12l}
&\text{~(i) \ \  with probability } \frac{1-y}2, 
\mathcal{P}(|\pmb{j} - My|\le 1 \big| |\tilde{y}\rangle)) \ge \frac8{\pi^2};\\
\label{eq12m}
&\text{(ii)\ with probability } \frac{1+y}2, 
\mathcal{P}(|\pmb{j}-M(1-y)|\le 1 \big| |\widetilde{1-y}\rangle) 
\ge \frac8{\pi^2},
\end{align}

where $\mathcal{P}(A\big| B)$ denotes the probability of an event $A$ 
conditioned on the event $B$.
\end{thm}
{\bf Proof:}  First, note from the definition (2.25) that 
\begin{align*}
|\tilde{y} \rangle  &= \mathcal{F}_M^{-1}(\frac{1}{\sqrt{M}}
\sum_{k=0}^{M-1} e^{i2k\pi y} |k \rangle )
=\frac{1}{\sqrt{M}}\sum_{k=0}^{M-1} e^{i2k\pi y}
\sum_{j=0}^{M-1}e^{-i2k\pi j/M}|j \rangle \\
&\equiv \sum_{k=0}^{M-1} \alpha_k(y)|k \rangle ,
\end{align*}
where
\[  \alpha_k(y) = \frac{1}{M} \sum_{j=0}^{M-1} e^{i2\pi j ( y  -\frac{k}{M})}.
\]
The probability that we will obtain $|\tilde{y}\rangle$ is $(1-y)/2$.  The
measurement of $|\tilde{y}\rangle$ will then yield an eigenstate $|k\rangle$ 
with 
probability $|\alpha_k(y)|^2$.  Our task now is to estimate $\alpha_k(y)$:
\begin{align}
|\alpha_k(y)| &=| \langle k|\tilde{y} \rangle | = 
| \langle k| \frac{1}{M} \sum_{p=0}^{M-1}
(\sum_{j=0}^{M-1}e^{i2\pi j(y-\frac{p}{M})})p \rangle |
\nonumber \\
&=\frac{1}{M} \Big|\sum_{j=0}^{M-1}e^{i2\pi j(y-\frac{k}{M})}\Big|
=\frac{1}{M}\Big| \frac{1-e^{i2\pi M(y-\frac{k}{M})}}
{1-e^{i2\pi(y-\frac{k}{M})}}\Big|
=\frac{1}{M}\Big|\frac{\sin(\pi(My-k))}{\sin(\pi(y-\frac{k}{M}))}\Big|.
\end{align}

We see in the above that $|\alpha_k(y)|^2$ is maximized if $y=k/M$, yielding
$|\alpha_k(y)|^2=1$, i.e., 
$\mathcal{P}(|k\rangle\textrm{ happens }\big| |\tilde{y}\rangle)=1$.  Thus,
the above provides a way of measuring $y$ in terms of $M$ and $k$.

In general, $y$ is a real number.  Therefore, we cannot expect the 
certainty $\mathcal{P}(|k\rangle\textrm{ happens }\big| |\tilde{y}\rangle)=1$ 
no matter how $M$ is 
chosen.  To treat the case $y \in \mathbb{R}$, we first define, for any
$r\in \mathbb{R}$,
\begin{align*}
\lfloor r \rfloor = \textrm{the largest integer smaller than } r,\\
\lceil r \rceil =\textrm{the smallest integer larger than } r.
\end{align*}

For fixed $M$, denote $\Delta=\frac{M y -\lfloor M y  \rfloor}{M}=
 y -\frac{\lfloor M y \rfloor}{M}$.  Then 
$\frac{1}{M}-\Delta=\frac{\lceil M y  \rceil-M y }{M}=
\frac{\lceil M y \rceil}{M}- y $.  Therefore, from (2.29),
\begin{align*}
\mathcal{P}(|My-k|\le 1\big| |\tilde{y} \rangle ) = 
\mathcal{P}(\lfloor My\rfloor =k\big| |\tilde{y} \rangle )+ 
\mathcal{P}(\lceil My\rceil=k \big| |\tilde{y} \rangle )\\
=\frac{\sin^2(M\Delta\pi)}{M^2\sin^2(\Delta\pi)}+
\frac{\sin^2(M(\frac{1}{M}-\Delta)\pi)}{M^2\sin^2(\frac{1}{M}-\Delta)\pi)};
\end{align*}

the RHS above attains minimum at $\Delta=\frac{1}{2M}$, giving
\begin{align*}
\mathcal{P}(|My-k|\le 1\big| |\tilde{y} \rangle ) &=
\frac{1}{M^2}(\frac{1}{\sin^2(\frac{\pi}{2M})}
+\frac{1}{\sin^2(\frac{\pi}{2M})})\\
&=\frac{2}{M^2\sin^2(\frac{\pi}{2M})} 
\ge \frac{2}{M^2(\frac{\pi}{2M})^2}
= \frac{8}{\pi^2}.
\end{align*}

Therefore (2.27) has been proven.

The second possibility is that, from (2.26), we obtain 
$|\widetilde{1-y} \rangle $ with
probability $\frac{1+y}{2}$; $|\widetilde{1-y} \rangle $ further collapses to 
$|k' \rangle $ such that
\begin{align*}
\mathcal{P}(|k'-M(1-y)|\le 1\big| |\widetilde{1-y} \rangle ) = 
\frac{\sin^2(M\Delta\pi)}{M^2\sin^2(\Delta\pi)}+
\frac{\sin^2(M(\frac{1}{M}-\Delta)\pi)}{M^2\sin^2(\frac{1}{M}-\Delta)\pi)}
\ge \frac{8}{\pi^2},
\end{align*}
where $\Delta \equiv \frac{M(1-y)-\lfloor M(1-y)\rfloor}{M}$.
$\hfill\square$

\begin{rem}
\begin{itemize}
\item[(i)] The quantum search procedures as culminated in (2.26) is 
\emph{hybrid} in the sense that it operates concurrently on continuous 
(i.e., $t$ ) and discrete (i.e., $m$ in QFT) variables (Lloyd [\ref{Ll}]).
\item[(ii)] In QC implementation, (assume that) qubits are used and, thus, 
$M=2^n$ for some positive integer $n$.  The circuit for estimating $y$ from
the ancilla register $|m\rangle$ (cf. (2.23)-(2.26)) may be found in Fig. 1.
$\hfill\square$
\end{itemize}
\end{rem}

\begin{center}
\epsfig{figure=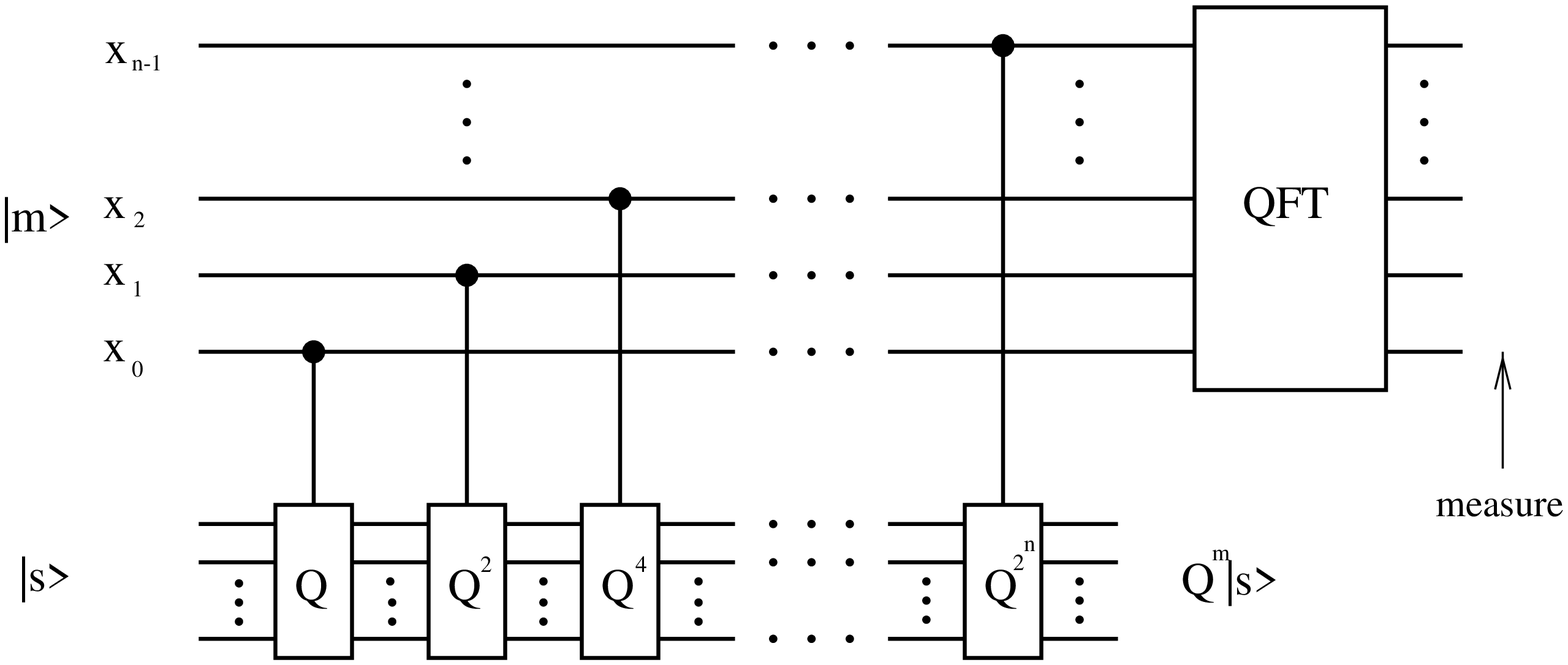,height=2.5in,width=5.5in}\\
\n {\bf Fig. 1 \ Circuit for estimating $\pmb{y}$ in (\ref{eq12k}), where
$\pmb{x_0,x_1,\ldots, x_{n-1}}$ represent the ascending order of qubits and
$\pmb{M=2^n}$.}
\end{center}

From (2.29), we see that in the estimation of $y$, what matters is 
$|sin(\pi(y-\frac{k}{M}))|$ and, consequently, the relevant distance between
our estimate $k/M$  and $y$ itself is not simply $|y-\frac{k}{M}|$.  
A better measurement of distance is given as follows.

\begin{defn}[{[\ref{Mo}, p. 45]}]  
The distance $d(y_1, y_2)$ between two real numbers $y_1$ and $y_2$ is the 
real number
\[
d(y_1, y_2) = \min_{j\in \mathbb{Z}} |y_1-y_2 +j|,
\]
i.e., $d(y_1, y_2)$ makes the shortest arclength on the unit circle
between $e^{i2\pi y_1}$ and $e^{i2\pi y_2}$ be $2\pi d(y_1, y_2)$.
$\hfill\square$
\end{defn}

\begin{cor}
Assume the same conditions as those in Thms. 2.1 and 2.4.  Measurement
of the first register of $|\Psi_2 \rangle $ on the RHS of (2.26) will yield the
state $|k \rangle $ such that 
\begin{itemize}
\item[(i)] if $My$ is an integer, then 
$\mathcal{P}(|k \rangle  \textrm{ happens })=1$;
\item[(ii)] if $My$ is not an integer, then
\begin{align}
\mathcal{P}(|k \rangle  \textrm{ happens }\big| |\tilde{y} \rangle )
=\frac{\sin^2(M\pi d(y, \frac{k}{M}))}{M^2\sin^2(\pi d(y, \frac{k}{M}))}
\le \frac{1}{(2M d(y, \frac{k}{M}))^2},\\
\mathcal{P}(|k \rangle  \textrm{ happens }\big| |\widetilde{1-y} \rangle )
=\frac{\sin^2(M\pi d(1-y, \frac{k}{M}))}{M^2\sin^2(\pi d(1-y, \frac{k}{M}))}
\le \frac{1}{(2M d(1-y, \frac{k}{M}))^2};
\end{align}
\item[(iii)] 
\begin{align*}
\mathcal{P}(d(y, \frac{k}{M})\le \frac{1}{M}\big| |\tilde{y} \rangle ) 
\ge \frac{8}{\pi^2},\\
\mathcal{P}(d(1-y, \frac{k}{M})\le \frac{1}{M}\big| |\widetilde{1-y} \rangle ) 
\ge \frac{8}{\pi^2};
\end{align*}
\item[(iv)] for $m>1$,
\begin{align}
\mathcal{P}(d(y, \frac{k}{M})\le \frac{m}{M}\big| |\tilde{y} \rangle ) \ge 
1-\frac{1}{2(m-1)};\\
\mathcal{P}(d(1-y,\frac{k}{M})\le \frac{1}{M}\big| |\widetilde{1-y}\rangle) \ge 
1-\frac{1}{2(m-1)}.
\end{align}
\end{itemize}
\end{cor}
{\bf Proof:}  Many estimates are already clear from the proofs given
previously.  The rest can be established using [\ref{Mo}, pp. 45-46] as follows.

It is clear from (2.29) that 
\begin{align}
\mathcal{P}(|k \rangle  \textrm{ happens }\big| |\tilde{y} \rangle )
=\frac{\sin^2(M\pi d(y, \frac{k}{M}))}{M^2\sin^2(\pi d(y, \frac{k}{M}))}.
\end{align}
Using the fact that $2x \le \sin \pi x \le \pi x$ for $x\in [ 0, 1/2]$,
from (2.34) we obtain
\[
\mathcal{P}(|k \rangle  \textrm{ happens }\big| |\tilde{y} \rangle ) \le
\frac{1}{M^2} \frac{1}{|2 d(y, \frac{k}{M})|^2},
\]
which proves (2.30).  We can similarly prove (2.31).

To show (2.32), we have
\begin{align*}
\mathcal{P}(d(y, \frac{k}{M}) \le \frac{m}{M}\big| |\tilde{y} \rangle )& =
\mathcal{P}(|My-k|\le m\big| |\tilde{y} \rangle )\\
&=1-\mathcal{P}(|My-k| >m\big| |\tilde{y} \rangle ) \\
&\ge 1-\sum_{j=m}^{M} \mathcal{P}(|My-k|=j\big| |\tilde{y} \rangle )\\
&\ge 1-\sum_{j=m}^{\infty}\mathcal{P}(|My-k|=j\big| |\tilde{y} \rangle )\\
&\ge 1-2\sum_{j=m}^{\infty}\frac{1}{4M^2(\frac{j}{M})^2}  
\ge 1-\frac{1}{2(m-1)}.
\end{align*}

The estimate (2.33) also follows similarly.
$\hfill\square$

\section{Efficiency and Complexity}
\setcounter{equation}{0}

Let us address various relevant issues in this section.
	
\vspace{0.2cm}
\noindent
\textbf{(I1)  Will the search algorithm with the availability of partial information
given in \S2 always be more efficient than the unstructured Grover 
multi-object search algorithm?}

The answer is NO.  A simple counterexample is sufficient to demonstrate
this point.  Let 
\begin{align} \label{3.1}
\mathcal{T} \subseteq A_1 \cup A_2,\   \mathcal{T} \subseteq A_1, \ 
\mathcal{T} \cap A_2 = \emptyset;\   A_1, A_2 \subseteq \bar{A}.
\end{align}

Assume that the cardinalities of, respectively, $\mathcal{T}, \bar{A},
A_1$, and $A_2$, are $l$, $N$, $n_1$ and $n_2$.  Let the reliability 
coefficients be $\{\alpha_1, \alpha_2\}$, where $\alpha_1, \alpha_2 >0$,
$\alpha_1+\alpha_2=1$.  Then by (3.1), we easily see that
\begin{align}
\left.\begin{array}{lll}
\beta_i &=&\alpha_1/\nu, i=1,2, \ldots, l;  \qquad \qquad
\textrm{(cf. (2.11), (2.12))}\\
\nu&=&[(n_1-n_{12})\alpha_1^2 +n_{12}(\alpha_1+\alpha_2)^2
+(n_2-n_{12})\alpha_2^2]^{1/2},\\
n_{12} &\equiv& \textrm{ the cardinality of } A_1 \cap A_2.
\end{array}\right\}
\end{align}

Thus
\begin{align}
y=\left(\sum_{i=1}^l \beta_i^2 \right)^{1/2}=
\left(\frac{l}{\nu^2}\alpha_1^2\right)^{1/2}
=\frac{\sqrt{l}}{\nu} \alpha_1 =\frac{\sqrt{l}}{\nu}(1-\alpha_2).
\end{align}
and by Cor. 2.2, the time $T$ required to reach $\mathcal{L}$ is
\begin{align}
T=\frac{\pi}{2Ey}= \frac{\pi \nu}{2E\sqrt{l}}\,\frac{1}{1-\alpha_2}.
\end{align}
If $\alpha_2$ is very close to 1, then it is easy 
to see from (3.2)--(3.4) that
\[  \lim_{\alpha_2 \to 1^-}  T = \infty.   \]

Therefore this algorithm is not efficient when $\alpha_2$ is close to $1$.
(Conversely, if $\alpha_2$ is close to $0^+$, then we see that the 
algorithm will be efficient.)

It is obvious to see what causes the trouble.  In (3.1), we see that
the information set $A_2$ is irrelevant to the search target set 
$\mathcal{T}$(i.e., $\mathcal{T}\cap A_2$) but too heavy weight $\alpha_2$
is assigned to the set $A_2$.  This is a situation with 
\emph{misplaced confidence} on the set $A_2$.  It is definitely to be avoided.  The opposite
situation of which is called by us one with \emph{basic confidence}.

\begin{defn}
Consider (2.2).  If $A_j\cap \mathcal{T} \ne \emptyset$ for $j=1,2, \ldots, n$,
then we say that we have basic confidence in the partial information sets
$A_1, A_2, \ldots, A_n$.
$\hfill\square$
\end{defn}
\noindent
\textbf{(I2) Will the search algorithm in \S 2, with the additional 
assumption of basic confidence, be more efficient than the 
unstructured Grover multi-object search algorithm?}

The answer is YES.  The following theorem shows that we still maintain a
quadratic speedup of Grover.

\begin{thm}
Consider (2.2) and assume that we have basic confidence.  Then we have
\begin{align}
y= \left(\sum_{i=1}^l \beta_i^2\right)^{1/2} \ge \frac{1}{ n^{1/2} (l+R)^{1/2}},
\end{align}
where $l+R$ is the totality of distinct elements in $A_1 \cup \ldots A_n$.
Consequently, the time $T$ required for $|\psi(T)\rangle$ to reach $\mathcal{L}$
is
\begin{align}
T=\frac{\pi}{2Ey} \le \frac{\pi n^{1/2}}{2E}(l+R)^{1/2}
\end{align}
\end{thm}

{\bf Proof:}  Comparing (2.11) and (2.12), we have, for each $j=1, \ldots, l$,
\[
\beta_j =\sum_{i=1}^n \frac{\alpha_{j,i}}{\nu},
\]
where \[
\alpha_{j,i} = \left\{\begin{array}{ll}
\alpha_i  & \textrm{if } |w_j\rangle \in A_i,\\
0 & \textrm{otherwise}.\end{array}\right.
\]
Therefore, 
\begin{align}
y^2 = \sum_{j=1}^l \beta_j^2 = 
\sum_{j=1}^l\left( \sum_{i=1}^n \frac{\alpha_{j,i}}{\nu}\right)^2 
\ge \frac{1}{\nu^2} \sum_{i=1}^n \alpha_i^2,
\end{align}
by the assumption that we have basic confidence and the inequality
$(a+b)^2 \ge a^2 +b^2$ if both $a$ and $b$ are positive. Also,
\begin{align}
\sum_{i=1}^n \alpha_i^2 \ge \frac{1}{n}
\end{align}
under the constraint that $\sum_{i=1}^n \alpha_i =1$.  This follows from
the Lagrange multiplier method (or the Cauchy-Schwarz inequality).

From (2.11), the normalization constant $\nu$ takes minimal value where
the sets $A_1, \ldots, A_n$ are in totality orthonormal, and takes maximal
value when it happens that $A_1 =A_2= \ldots =A_n =\{|w_j\rangle \big| 
j=1, 2, \ldots, l+R\}$.  Thus
\begin{align}
\sum_{j=1}^n\sum_{i=1}^{k_j} \alpha_j^2 \le \nu^2 \le l+R.
\end{align}

By (3.8) and (3.9), we obtain
\[ 
y^2 \ge \frac{1}{\nu^2} \sum_{i=1}^n\alpha_i^2 \ge \frac{1}{(l+R)n},
\quad \textrm{ i.e., (3.5)},
\]
and hence (3.6).
$\hfill\square$

\begin{cor}
Assume basic confidence.  If $l+R= O(N^\delta)$ for some small $\delta >0$, 
then the search task for the first item will be completed in time duration
$T=O(n^{1/2}N^{\delta/2})$, where $n$ is the cardinality of the set
$\{A_1, \ldots, A_n\}$.
$\hfill\square$
\end{cor}

Normally, if the partial information sets are very descriptive in the sense
that $l+R$ is small, say, $l+R =O(N^\delta)$ with $\delta \ll 1$, then the
search algorithm  in \S 2 will be more efficient than the unstructured 
Grover search.

\begin{rem}
\begin{itemize}
\item[(i)] The estimate (3.6) is obtained under the possibility that
$A_1=A_2=\ldots=A_n=\{|w_j\rangle\big| j=1, 2, \ldots, l+R\}$, which is a rare
and trivial happenstance (that all information sets coincide).  The other
extreme is that there is no overlapping at all between the information sets,
i.e., $A_i\cap A_j =\emptyset$ for any $i, j \in \{1,2, \ldots, n\}, i\ne j$.
Then under the assumption of basic confidence the conclusion in Cor. 3.2
still maintains its order of optimality .  See (ii) and Cor. 3.3 below.
\item[(ii)] By observing (2.11) and (2.12), we see that for the example
(1.1) and (1.2),  any $w_{j_0} \in \mathcal{T}$ such that $w_{j_0} \in 
A_1\cap A_2 \cap A_3$ will have a larger weight $\beta_{j_0}$ because
$w_{j_0}$ is repeated in all $A_1, A_2$ and $A_3$.  As a consequence,
this $w_{j_0}$ is likely to be the outcome as the search of the first item.
This means that a book title including \emph{all} the interests in hunting, 
fishing
and hiking is more likely to turn up than the other titles as the outcome
of search.  This can be undesirable, however.  The only way to avoid this 
from happening is to eliminate the repetitions (or overlappings) between
all $A_1, A_2$ and $A_3$ (and, in general, between all $A_1, A_2, \ldots,
A_n$).  Indeed, under the assumption that $A_i \cap A_j = \emptyset$ for
all $i, j \in \{1, \ldots, n\}$ and $i\ne j$, we have (from (2.11))
\begin{align}
\nu^2 &= \sum_{j=1}^{n}\sum_{i=1}^{k_j} \alpha_j^2 
=k_1\alpha_1^2+k_2\alpha_2^2+\ldots+k_n\alpha_n^2 \nonumber\\
& \le (k_1+k_2+\ldots+k_n)(\alpha_1^2 +\alpha_2^2 +\ldots +\alpha_n^2) 
\nonumber\\
& =(l+R) \sum_{i=1}^n \alpha_i^2.
\end{align}
Using (3.10) in (3.7), we obtain
\[
y^2 \ge \frac{1}{l+R}
\]
and hence \[ T \le \frac{\pi}{2E} (l+R)^{1/2}.\]
$\hfill\square$
\end{itemize}
\end{rem}

\begin{cor}
Assume basic confidence and that $A_i\cap A_j =\emptyset$ for all 
$i, j \in  \{1,2, \ldots, n\}, i\ne j$.  Then
\[ 
y\ge \frac{1}{(l+R)^{1/2}}, \ 
T=\frac{\pi}{2Ey} \le \frac{\pi}{2E} (l+R)^{1/2}.
\]
Consequently, if $l+R= O(N^\delta)$, then the search task for the first item
will be completed in time duration $T=O(N^{\delta/2)}$ independent of $n$.
$\hfill\square$
\end{cor}
\noindent
\textbf{(I3) Can we determine $l$, the cardinality of $\mathcal{T}$, using the 
algorithm in \S 2?}

The answer is NO, unless we do extra work.  In general, because the choice
of reliability coefficients $\{\alpha_j\}_{j=1}^n$ is somewhat arbitrary, the
cardinality $l$ of $\mathcal{T}$ will not be manifested in $y$.  Even if we
choose uniform weights $\alpha_i =1/n, i=1,2, \ldots, n$, for the 
information sets $A_1, \ldots, A_n$, we still are unable to estimate $y$
(or $1-y$) because elements in $\mathcal{T}$ may have repeated appearances
in $A_1, A_2, \ldots, A_n$.  When all the $A_i$'s are disjoint, then
\[
y=\left(\frac{l}{l+R}\right)^{1/2}.
\]
One can thus estimate $l$ and $R$ based on $y$ and $1-y$ from Cor. 2.5,
as it is usually done in solving the \emph{quantum counting problem}.

Because the information sets $A_1, A_2, \ldots, A_n$ generally have some
overlapping, \emph{we need to eliminate such overlapping} first through some 
processing in order to do counting.

\vspace{0.2cm}
\noindent
\textbf{(I4)  The choice of different information sets}

The example stated in (1.1) so far has been treated by choosing the
information sets $A_1$, $A_2$ and $A_3$ as denoted in the paragraph
following (1.2).  Instead, one can choose just a single information
set
\[ 
A_0=\{\textrm{book title } x \big| x \textrm{ is published between }
1/15/1990 \textrm{ and } 6/15/1990\}.
\]
Then the search of $\mathcal{T}$ will be carried out in $A_0$.  As we 
expect the cardinality of $A_0$ will be much larger than the sum of the
cardinalities of $A_1, A_2$ and $A_3$, this search will be less efficient.

The choice of information sets seems to rely on the human operator as
well as on how the data are encoded.

\vspace{2mm}
\noindent
\textbf{(I5) The work involved}

For each estimation of $y$, we need $O((\log{M})^2)$ number of operations.  With
each (estimated) value of $y$, we require time duration $T=\frac{2\pi}{Ey}$
in order to obtain the first search item.

\end{document}